\documentclass{aastex}
\usepackage{spr-astr-addons}
\usepackage{url}

\RequirePackage{color}

\newcommand{\emaila}{jameel@giki.edu.pk}
\usepackage{latexsym}
\usepackage{graphics}
\usepackage{graphicx}
\usepackage{epsfig}

\begin{document}

\title{Gamow-Teller strength distribution in
proton-rich nucleus $^{57}$Zn and its implications in astrophysics}
\shortauthors{Nabi and Rahman} \shorttitle{GT strength in
proton-rich $^{57}$Zn}
\author{Jameel-Un Nabi \altaffilmark{1}}
\affil{Faculty of Engineering Sciences, GIK Institute of Engineering
Sciences and Technology, Topi 23640, Swabi, Khyber Pakhtunkhwa,
Pakistan} \email{\emaila} \and \author{Muneeb-Ur Rahman} \affil{
Department of Physics, Kohat University of Science and Technology,
Kohat 26000, Khyber Pakhtunkhwa, Pakistan}
\altaffiltext{1}{Corresponding author email : jameel@giki.edu.pk}
\begin{abstract}
Gamow-Teller (GT) transitions play a preeminent role in the collapse
of stellar core in the stages leading to a Type-II supernova. The
microscopically calculated GT strength distributions from ground and
excited states are used for the calculation of weak decay rates for
the core-collapse supernova dynamics and for probing the concomitant
nucleosynthesis problem. The B(GT) strength for $^{57}$Zn is
calculated in the domain of proton-neutron Quasiparticle Random
Phase Approximation (pn-QRPA) theory. No experimental insertions
were made (as usually made in other pn-QRPA calculations of B(GT)
strength function) to check the performance of the model for
proton-rich nuclei. The calculated B(GT) strength distribution is in
good agreement with measurements and shows differences with the
earlier reported shell model calculation. The pn-QRPA model
reproduced the measured low-lying strength for $^{57}$Zn better in
comparison to the KB3G interaction used in the large-scale shell
model calculation. The stellar weak rates are sensitive to the
location and structure of these low-lying states in daughter
$^{57}$Cu. The structure of $^{57}$Cu plays a sumptuous role in the
nucleosynthesis of proton-rich nuclei. The primary mechanism for
producing such nuclei is the rp-process and is believed to be
important in the dynamics of the collapsing supermassive stars.
Small changes in the binding and excitation energies can lead to
significant modifications of the predictions for the synthesis of
proton rich isotopes. The $\beta^{+}$-decay and electron capture
(EC) rates on $^{57}$Zn are compared to the seminal work of Fuller,
Fowler and Newman (FFN). The pn-QRPA calculated $\beta^{+}$-decay
rates are generally in good agreement with the FFN calculation.
However at high stellar temperatures the calculated
$\beta^{+}$-decay rates are almost half of FFN rates. On the other
hand, for rp-process conditions, the calculated electron capture
($\beta^{+}$-decay) rates are bigger than FFN rates by more than a
factor 2 (1.5)  and may have interesting astrophysical consequences.
\end{abstract}

\section{Introduction}
\label{intro} The weak interactions and gravity are the guru that
drive the evolution of heavy mass stars and the concomitant
nucleosynthesis, which has been the subject of much computation. The
incarnation of the core-collapse mechanism is the conversion of a
fraction of the gravitational energy into kinetic energy of the
ejecta and internal energy of the inner core of the exploding star
\cite{Bur06}. This inner core is mainly composed of iron group
nuclei and when this inner core exceeds the appropriate
Chandrasekhar mass limits it becomes unstable and inaugurates the
implosion of the inner core and consequently announces the death of
the star in a catastrophic explosion. The relic is either a neutron
star or black hole depending on the mass of the progenitors. If the
stars are not too massive then they collapse and consequently bounce
and explode in spectacular visual display commonly known as type-II
or Ib/c supernovae. The collapse of the star is very sensitive to
the entropy and to the number of leptons per baryon, $Y_{e}$
\cite{Nabi05,Bethe79}. The neutrinos are considered as main sink of
energy and lepton number until the core density reaches around
$10^{10}g-cm^{-3}$. At later stage of the collapse this assumption
is no longer legitimate as weak interaction rates increase with
increase of stellar core density. For densities $> 10^{11}
g-cm{^-3}$, the neutrinos mean free paths become shorter and
consequently they proceed through all phases of free streaming,
diffusion, and trapping.  The most tightly bound of all the nuclei
in the inner core of star is $^{56}$Fe \cite{Cowan04} and further
fusing of the nuclei is highly endothermic. The second bottleneck to
the synthesis of heavier elements is the high Z number which poses
high coulombic barrier for the charged particles to initiate nuclear
reactions at stellar core temperatures. This impediment to further
nucleosynthesis is rescinded with the help of neuron capture
processes in the stellar core to form heavier isotopes beyond iron
group nuclei. These processes are further classified into slow (s-)
and rapid (p-) neutron capture processes depending on the neutron
capture time scale $\tau_n$ and beta decay time scale $\tau_\beta$
for nucleus to endure beta decay. As discussed earlier that weak
interaction and gravity are the mentor that drives the stellar
evolutionary process and its subsequent death in a cataclysmic
explosion. The main weak interaction processes that play an
effective role in the stellar evolution are beta decays, electron
and positron capture processes and neutrinos emission/capture
processes subject to the physical conditions available for these
processes in the stellar core. These weak decay processes are
smitten by Fermi and Gamow-Teller (GT) transitions. Fermi
transitions are straightforward and are important only for beta
decays. For nuclei with $N > Z$, Fermi transitions are Pauli blocked
and the GT transitions dominate and their calculation is model
dependent. Spin-isospin-flip excitations in nuclei at vanishing
momentum transfer are commonly known as GT transitions. These
transitions are ideal probe to test nuclear structure and play
preeminent role in the nucleosynthetic origin of the elements in the
late phases of the stellar life. When the stellar matter is
degenerate then the phase space for the electron in beta decay is
Pauli blocked and electron captures become dominant in the stellar
core producing neutrinos which escape from the surface of the star
and takes away the core energy and entropy as well.  These electron
capture rates and $\beta^{+}$ decay rates are very sensitive to the
distribution of the $GT_{+ }$ strength which is responsible for
changing a proton into a neutron (the plus sign is for the isospin
raising operator ($t_{+}$), present in the GT matrix elements, which
converts a proton into a neutron). The authors in
\cite{Wanajo06,Kitaura06} took the accurate neutrino transport into
account in their hydrodynamic studies of core-collapse supernovae
and showed that the bulk of the neutrino-heated ejecta is
proton-rich during the early phase $(\leq1 s)$. Their studies also
provided that it is likely possible that the neutrino-induced
rp-process takes place in all core-collapse supernovae and other
astrophysical sites such as collapsar jets or disk winds formed
around a black hole. In reference \cite{Soderberg08} the authors
discovered an extremely luminous X-ray outburst that marked the
birth of a supernova of Type Ibc. They attributed the X-ray outburst
to the break-out of the supernova shock-wave from the progenitor.

The open shell nuclei with a few nucleons outside a doubly magic
shell closure are of colossal interest to test the nuclear model
predictions. $^{57}$Cu is of paramount importance in this regard to
test the pn-QRPA predictions. The \textit{pf}-shell nucleus
$^{57}$Cu has a single proton just above the closed core of
even-even $^{56}$Ni with Z = N = 28. This simple structure permits
far more accurate model calculations than are possible in the middle
of a nuclear shell. In particular, a comparison of the low-lying
levels of $^{57}$Cu with the well-determined excited states of its
mirror nucleus $^{57}$Ni is important for studying the charge
symmetry of the nucleus. The doubly magic nature of $^{56}$Ni
confers it a stable structure and elements beyond $^{56}$Ni are
cooked in the stellar pot only via $^{56}$Ni(p, $\gamma $)$^{57}$Cu
reaction. The authors in Ref. \cite{Schatz98} pointed out that the
structure of $^{57}$Cu plays a sumptuous role in the nucleosynthesis
of proton-rich nuclei. This points the fact that this reaction rate
is susceptible to the structure of $^{57}$Cu and environs nuclei,
including their binding energies. The primary mechanism for
producing such nuclei is the rp-process and is believed to be
important in the dynamics of the collapsing supermassive stars and
x-ray bursts \cite{Wallace81}. The rp-process is characterized by
proton capture reaction rates that are orders of magnitude faster
than any other competing process, specially $\beta$-decay rates. The
reaction path follows a series of fast (p,$\gamma$)-reactions until
further proton capture is inhibited, either by negative proton
capture Q-values (proton decay) or small positive proton capture
Q-values (photodisintegration). As electrons captures in the stellar
pot assist the cooking process and gravity, therefore, it is crucial
to know the nuclear structure properties of the doubly magic shell
$^{56}$Ni and nuclei in its vicinity. These nuclei drive the
electron capture processes in the dense core of heavy mass stars.
The beta decay rate calculations require the evaluation of the GT
strengths for many levels per nucleus. The pn-QRPA theory gives us
the emancipation to use model space as big as $7\, \hbar \omega $
rather than truncated model spaces usually employed in some shell
model calculations.

Earlier the half-lives for $\beta^{+}$/EC decays for
neutron-deficient nuclei with atomic numbers Z = 10 - 108 were
calculated up to the proton drip line for more than 2000 nuclei
using the same model \cite{Hir93}. These microscopic calculations
gave a remarkably good agreement with the then existing experimental
data (within a factor of two for more than 73$\%$ of nuclei with
experimental half-lives shorter than 1 s for $\beta^{+}$/EC decays).
Most nuclei of interest of astrophysical importance are the ones far
from stability and one has to rely on theoretical models to estimate
their beta decay properties. The accuracy of the pn-QRPA model
increases with increasing distance from the $\beta$-stability line
(shorter half-lives) \cite{Hir93,Sta90}. This is a promising feature
with respect to the prediction of experimentally unknown half-lives
(specially those present in the stellar interior), implying that the
predictions are made on the basis of a realistic physical model.

Nucleosynthesis in proton-rich ejecta occur at low densities. At
high densities the composition is neutron-rich. Under terrestrial
conditions $^{57}$Zn $\beta^{+}$ decays to $^{57}$Cu. We used the
calculated ground and excited state GT strength functions to
microscopically calculate the $\beta^{+}$/EC rate of $^{57}$Zn in
stellar matter. In proton-rich supernova environments $^{57}$Zn is
produced at high temperatures where nuclear statistical equilibrium
is valid. According to studies by authors in Ref. \cite{Schatz98}
the peak conditions for rp-process are in the vicinity of T = (1--3)
GK and $\rho = (10^{6}-10^{7}$) gcm$^{-3}$. The current analysis
shows that the electron capture and positron decay rates of
$^{57}$Zn contribute roughly equally to the total weak rates at peak
rp-process conditions (with the positron decay rates bigger roughly
by a factor of 6--7). At still higher temperatures and densities the
electron capture on $^{57}$Zn dominates . As a result both
$\beta^{+}$ and electron capture stellar rates on $^{57}$Zn are
being presented in this paper.

Section 2 describes the theoretical formalism of pn-QRPA
calculation. Section 3 discusses the pn-QRPA calculated GT strength
functions and its comparison with measurements and previous
calculation. The calculated weak rates are presented in Section 4.
Here we also compare our results with the pioneering calculation of
Fuller and collaborators \cite{Fuller82}. We finally conclude our
findings in Section 5.

\section{Model Description}
The nuclei involved in the stellar interior have finite probability
of occupation of excited states and the state by state evaluation of
the GT strength distribution from these excited states is a
formidable task. The pn-QRPA is considered an efficient model to
extract the GT strengths for the ground as well as excited states of
the involved nuclei in stellar matter thanks to the huge available
model space of seven major shells. The transitions from the excited
states contribute effectively to the total electron capture rate and
a microscopic calculation of excited state GT strength distributions
is desirable. The electron capture rates play a pivotal role in the
dynamics of core collapse of stars. The pn-QRPA model is used in the
present work to calculate the GT strength functions and associated
electron capture/$\beta^{+}$-decay rates for proton-rich nucleus
$^{57}$Zn.

The Hamiltonian of the pn-QRPA is given by
\begin{equation} \label{GrindEQ__1_}
{\rm H}^{{\rm QRPA}} {\rm \; =\; H}^{{\rm sp}} {\rm \; +\; V}^{{\rm
pair}} {\rm \; +\; V}_{{\rm GT}}^{{\rm ph}} {\rm \; +\; V}_{{\rm
GT}}^{{\rm pp}},
\end{equation}
where $H^{sp} $ is the single-particle Hamiltonian, $V^{pair} $ is
the pairing force, $V_{GT}^{ph} $  is the particle-hole (ph) GT
force, and $V_{GT}^{pp} $ is the particle-particle (pp) GT force.
Pairing is treated in the \emph{BCS} approximation, where an assumed
constant pairing force with force strength $G$ ($G_{p}$ and $G_{n}$
for protons and neutrons, respectively) is applied,
\begin{eqnarray} \label{GrindEQ__2_}
V^{pair} =-G\sum _{jmj'm'}(-1)^{l+j-m}  c_{jm}^{+} c_{j-m}^{+}
(-1)^{l'+j'-m'} \nonumber \\ c_{j'-m'} c_{j'm'},
\end{eqnarray}
where the sum over $m$  and $m'$  is restricted to $m$, ${m'}> 0$,
and $l$ donates the orbital angular momentum.

In the present work, in addition to the well known particle-hole GT
force \cite{Halbleib67,Staudt90,Muto92}
\begin{equation} \label{GrindEQ__3_}
V_{GT}^{ph} =2\chi \sum _{\mu }(-1)^{\mu } Y_{\mu }  Y_{-\mu }^{+},
\end{equation}
with
\begin{equation} \label{GrindEQ__4_}
Y_{\mu } =\sum _{j_{p} m_{p} j_{n} m_{n} }<j_{p} m_{p} |t{}_ {-}
\sigma _{\mu }  |j_{n} m_{n} >c_{j_{p} m_{p} }^{+} c_{j_{n} m_{n} },
\end{equation}
the particle-particle GT force \cite{Soloviev87,Kuzmin88}
\begin{equation} \label{GrindEQ__5_}
V_{GT}^{pp} =-2\kappa \sum _{\mu }(-1)^{\mu }  P_{\mu }^{+} P_{-\mu
},
\end{equation}
with
\begin{eqnarray} \label{GrindEQ__6_}
P_{\mu }^{+} =\sum _{j_{p} m_{p} j_{n} m_{n} }<j_{n} m_{n} |(t{}_{-}
\sigma _{\mu } )^{+} |j_{p} m_{p} >(-1)^{l_{n} +j_{n} -m_{n} }
\nonumber \\ c_{j_{p} m_{p} }^{+} c_{j_{n} -m_{n} }^{+},
\end{eqnarray}

is also taken into account.

The capture/decay rate of a transition from the \textit{i}th state
of a parent nucleus (Z, N) to the $jth$ state of the daughter
nucleus $(Z - 1, N + 1)$ is given by

\begin{equation} \label{GrindEQ__7_}
\lambda _{ij} \, =\, \ln 2\frac{f_{ij} (T,\rho ,E_{f} )}{(ft)_{ij}
},
\end{equation}

$f_{ij} $ is the phase space integral. The $(ft)_{ij} $ of an
ordinary $\beta ^{\pm } $ decay from the state ${\left| i
\right\rangle} $ of the mother nucleus to the state ${\left| f
\right\rangle} $ of the daughter is related to the reduced
transition probability $B_{ij} $ of the nuclear transition by

\begin{equation} \label{GrindEQ__8_}
(ft)_{ij} \, =\, {D\mathord{\left/ {\vphantom {D B_{ij} .}} \right.
\kern-\nulldelimiterspace} B_{ij}.}
\end{equation}

The D appearing in Eq. 8 is compound expression of physical
constants,

\begin{equation} \label{GrindEQ__9_}
D\, =\, \frac{2\pi ^{3} \hbar ^{7} \ln 2}{g_{V}^{2} m_{e}^{5} c^{4}
},
\end{equation}

and the reduced transition probability of the nuclear transition is

\begin{equation} \label{GrindEQ__10_}
B_{ij} \, =\, B(F)_{ij} \, +\, \left({g_{A} \mathord{\left/
{\vphantom {g_{A}  g_{V} }} \right. \kern-\nulldelimiterspace} g_{V}
} \right)^{2} B(GT)_{ij},
\end{equation}

The value of D = 6146 $\pm$ 6 s \cite{Jokinen02} is adopted and the
ratio of the axial-vector $(g_{A} )$ to the vector $(g_{V} )$
coupling constant is taken as -1.257. $B(F)_{ij} $ and $B(GT)_{ij} $
are reduced transition probabilities of the Fermi and GT
transitions, respectively. These reduced transition probabilities of
the nuclear transition are given by,

\begin{equation} \label{GrindEQ__11_}
B(F)_{ij} \, =\, \frac{1}{2J_{i} +1} |<j||\sum _{k}t_{\pm }^{k} ||\,
i>|^{2},
\end{equation}

\begin{equation} \label{GrindEQ__12_} B(GT)_{ij} \, =\, \frac{1}{2J_{i} +1}
|<j||\sum _{k}t_{\pm }^{k} \vec{\sigma }^{k} ||\,  i>|^{2}.
\end{equation}

Calculation of phase space integrals can be seen from Ref.
\cite{Nabi99}.

The number density of electrons associated with protons and nuclei
is $\rho Y_{e} N_{A} ,$ where $\rho $ is the baryon density, $Y_{e}
$ is the ratio of electron number to the baryon number, and $N_{A} $
is the Avogadro's number.

\begin{equation} \label{GrindEQ__14_}
\rho Y_{e} \, =\, \frac{1}{\pi ^{2} N_{A} } (\frac{m_{e} c}{\hbar }
)^{3} \int _{0}^{\infty }(G_{-}  -G_{+} )p^{2} dp,
\end{equation}

where $p\, =\, (w^{2} -1)^{1/2} $ is the electron or positron
momentum.

The total capture/decay rate per unit time for a nucleus in thermal
equilibrium at temperature $T$ for any weak process is then given by

\begin{equation} \label{GrindEQ__15_}
\lambda  \, =\, \sum _{ij}P_{i} \lambda _{ij}.
\end{equation}

Here $P_{i} $ is the probability of occupation of parent excited
states and follows the normal Boltzmann distribution. The summation
over all initial and final states is carried out until satisfactory
convergence in the rate calculations is achieved.

\section{Gamow-Teller Strength Distributions}
The charge exchange reactions with high resolution are important
probe for the study of nuclear structure in astrophysics. The GT
strengths play an important role in electron capture and beta decay
processes in the dynamics of stellar collapse
\cite{Nabi07,Fujita05}. Experimentally the (p, n), (n, p), (d, 2He),
and (2He, t) reactions can be used to probe the GT transitions (in
both directions) at higher excitation energies. Turning to theory
one sees that a large amount of calculations of weak interaction
rates for astrophysical applications have become available in recent
times (e.g. \cite{Nabi99,Heger01,Langanke01,Rahman07,Nabi04}).

\begin{figure}[t]
\includegraphics[width=3.3in,height=4.3in]{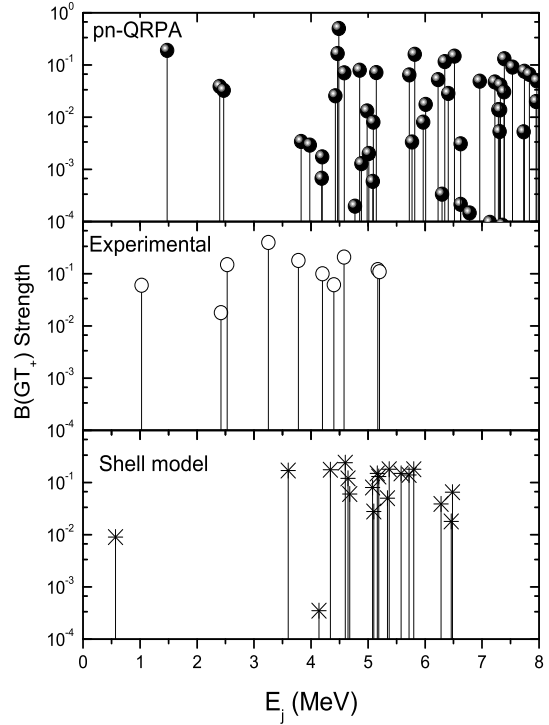}
\caption {The comparison of  pn-QRPA, experimental, and shell model
B(GT$_{+}$) strength distribution of $^{57}$Zn  as a function of
$^{57}$Cu excitation energy.}\label{fig1}
\end{figure}

To allow the reader the judgment of quality achieved in the pn-QRPA
model underlying the stellar weak interaction rate calculation, the
B(GT) strength calculated using the pn-QRPA theory is compared with
experimental \cite{Jokinen02} and with B(GT) strengths predictions
obtained from a large-scale shell model calculation using the
effective interaction KB3G \cite{Poves01}. The results are depicted
in Fig.~\ref{fig1}. Vieira et al. \cite{Vieira76} for the first time
probed the structure of $^{57}$Cu in the beta decay of $^{57}$Zn by
employing the $^{40}$Ca($^{20}$Ne, 3n) fusion evaporation reaction
and by Zhou et al. \cite{Zhou96} in the $^{1}$H($^{58}$Ni,
$^{57}$Cu-$\gamma $)2n reaction by using the recoil mass
spectrometer MARS at the Texas A\&M Cyclotron Institute.
Considerable improvements were obtained in Ref. \cite{Jokinen02}
concerning the quality of the experimental data, in particular with
respect to source purity and proton energy resolution.

\begin{figure}[t]
\includegraphics[width=3.2in,height=4.3in]{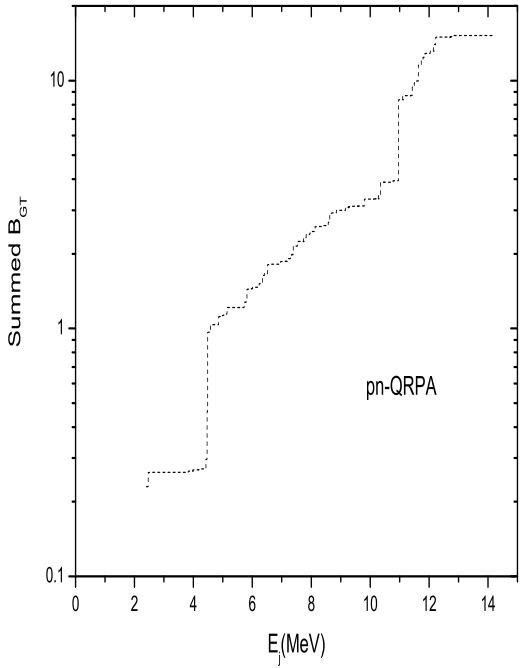}
\caption {The pn-QRPA calculated integrated strength within the
$Q_{EC}$ window. $E_{j}$  represents the energy in daughter nucleus
$^{57}$Cu in units of MeV.}\label{fig2}
\end{figure}
The authors in Ref. \cite{Jokinen02} investigated the beta-delayed
proton decay of $^{57}$Zn at GSI online isotope separator. $^{57}$Zn
nuclei were produced in fusion evaporation reactions,
$^{28}$Si($^{32}$S, 3n), by using a 150 MeV $^{32}$S beam on a
$^{28}$Si target and then the beta-delayed protons were measured
with high resolution by employing a charged-particle detector. The
GT strength distributions observed in this experiment is shown in
the middle panel of Fig.~\ref{fig1}. The morphology of the pn-QRPA
GT strength (upper panel) for beta transitions between the ground
state of $^{57}$Zn to the ground state of daughter $^{57}$Cu is in
good agreement with the measured data of Ref. \cite{Jokinen02}.
These transitions are of allowed nature in daughter $^{57}$Cu. The
experimental GT strength between 2 and 3 MeV is well reproduced by
pn-QRPA. These peaks were missing in the large-scale shell model
calculation as shown in the bottom panel of Fig.~\ref{fig1}. We used
the $Q$-value of 14.51 MeV for $^{57}$Zn from the recent
compilations of Ref. \cite{Audi03a,Audi03b}.

\begin{figure}[t]
\includegraphics[width=4.3in,height=4.3in]{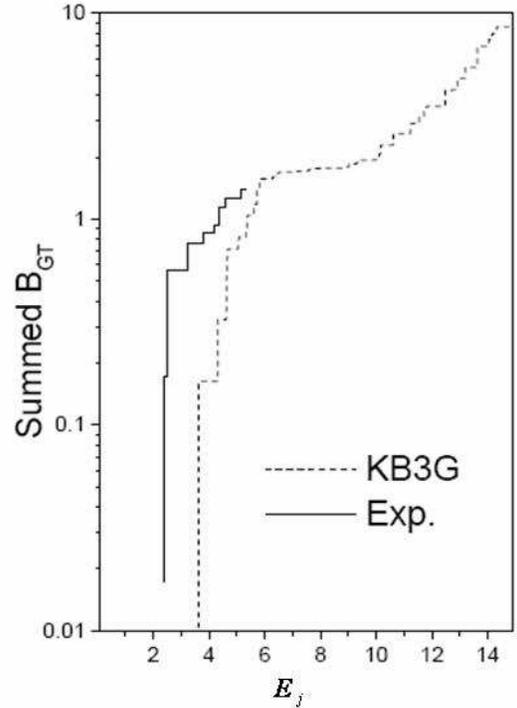} \caption {Comparison of
experimental and shell model integrated GT strength within the
$Q_{EC}$ window. The figure is reproduced from Ref.
\cite{Jokinen02}.}\label{fig3}
\end{figure}
It is evident that the GT strength distribution calculated by the
shell model KB3G, as compared to pn-QRPA and experimental results,
tends to shift to higher excitation energies in daughter $^{57}$Cu.
The pn-QRPA integrated GT strength over the $Q$ window is shown in
Fig.~\ref{fig2}. For comparison, the overall shift in B(GT) strength
of the shell model prediction along with the experimentally
extracted integrated GT strength is shown in Fig.~\ref{fig3}. In the
present work we obtained a summed B(GT) of 1.86 as compared to 1.25
in Ref. \cite{Jokinen02} within the excitation energy interval
between 0 and 7 MeV in daughter $^{57}$Cu. The authors in Ref.
\cite{Jokinen02} increased the strength by the upper B(GT) limit for
the 1.028 MeV state and the theoretical B(GT) value for the IAS to
yield $1.44^{+0.25}_{-0.17}$ and they estimated that less than 2\%
of the total GT strength could be missed due to the unobserved
gamma-transitions. The shell model predicted strength for the same
excitation of energies amounts to 1.68.

In beta decay experiments the sensitivity of the experiment strongly
decreases as one moves to higher excitation energies in daughter.
This puts a limit for such experiments and thus one can only probe
the low-energy tails of the GT strength distribution. Consequently
one has to rely on theoretical models for the calculation of GT
strength distributions in high excitation energies region as well as
for parent excited states. Within the $Q$ window of the reaction, we
extracted a total GT strength of 15.23 and the KB3G calculation
amount it to be 8.63. The pn-QRPA results include a quenching factor
of 0.8 as usually employed for the \textit{pf}-shell nuclei. The
luxurious $7\, \hbar \omega $ model space of the pn-QRPA facilitates
one to extract the state by state evaluation of the these GT
strength for higher excited states as well without assuming the
Brink's hypothesis and model space truncation as usually employed in
the shell model calculations. The truncation level, i.e, the number
of nucleons which are to be excited from the $f_{7/2}$ orbital to
the rest of the \textit{pf}-shell, in the KB3G calculation was
restricted to 4 nucleons for parent $^{57}$Zn and 5 nucleons for
daughter $^{57}$Cu. The pn-QRPA model extracted a bigger total
strength as compared to the KB3G interactions employed in the shell
model calculation. This certainly affects the calculation of
$\beta^{+}$-decay and electron capture rates in the stellar core at
high densities. Except at higher temperatures, the double shell
closure of $^{56}$Ni confers it stability compared to any of its
neighbors and the main path of the rp-process passes through
$^{56}$Ni. These properties pointed to the fact that heavier nuclei
may be produced by the rate of the radiative proton capture reaction
$^{56}$Ni(p, $\gamma $)$^{57}$Cu and confirm the candidature of
$^{56}$Ni as a waiting point nucleus in the reaction network. The
rate of this reaction in the stellar kilns is due almost entirely to
the first four resonances \cite{Zhou96} and particularly very
sensitive to the location and structure of the low-lying states in
daughter $^{57}$Cu. The pn-QRPA accounted well for these low-lying
states in $^{57}$Cu. The proton separation energy of $^{57}$Cu is
merely \textit{Sp} = 694.87 keV \cite{Audi03a,Audi03b}, and so all
of its excited states are resonances in this reaction. The authors
in Ref. \cite{Zhou96} stressed that small changes in the binding and
excitation energies lead to significant modifications of the
predictions for the synthesis of proton rich isotopes with A $>$ 56
and possibly for the time evolution of cosmic x-ray bursts.

We calculated the ground state and excited states GT strength
functions of $^{57}$Zn using the pn-QRPA model. The calculation was
performed for a total of 259 excited states in $^{57}$Zn covering an
excitation energy of around 30 MeV. The density of states were
chosen accordingly and contributions to GT strength from all excited
states were incorporated in the rate calculation. The ASCII files of
these GT strength functions are available and can be requested from
the corresponding author.

\section{$\beta^{+}$-Decay and Electron Capture Rates}
At low stellar densities and temperatures, $\beta^{+}$-decay is the
dominant mode for $^{57}$Zn to transform to $^{57}$Cu.  For the
calculation of the stellar weak interaction rate under corresponding
physical conditions, it is of colossal importance to reproduce the
measured GT strength distribution as compared to the total GT
strength due to very strong energy dependence of the phase space
factors. As discussed earlier the pn-QRPA model reproduced the
low-lying strength for $^{57}$Zn better in comparison to the KB3G
interaction used in the large shell model calculation. The stellar
weak rates are sensitive to the location and structure of these
low-lying states in daughter $^{57}$Cu. At the lowest temperature
considered in this work (T$_{9} [K] = 0.01$), excited parent states
are not appreciably populated (where T$_{9}$ is the stellar
temperature in units of $10^{9} K$) , while at low density ($\rho
Y_{e}$ = 10 gcm$^{-3}$), the continuum electron density is quite low
and the stellar rates should be close to the terrestrial values. At
this value of temperature and density the calculated half-life for
$\beta^{+}$-decay on $^{57}$Zn is 39.6 ms which is in excellent
agreement with the measured value of 40 ms \cite{Vieira76}.

\begin{table}[t]
\tiny \caption{$\beta^{+}$-decay rates (in units of $sec^{-1}$) and
ratio of $\beta^{+}$-decay to electron capture rates on $^{57}$Zn
for different selected densities and temperatures in stellar matter.
$\rho Y_{e}$ denotes the stellar density in units of $g/cm^{3}$ and
T$_{9}$ represents the temperature in $10^{9}$ K.} \label{tab1}
\begin{tabular}{llll|llll}
\\ \hline $\rho Y_{e} $ & $T_{9}$  &
$\lambda_{\beta^{+}} $ & R($\beta^{+}/EC$) &
$\rho Y_{e} $ & $T_{9}$ & $\lambda_{\beta^{+}} $ & R($\beta^{+}/EC$)\\
\noalign{\smallskip}\hline\noalign{\smallskip}
10 & 0.01 & 17.50 & 5.6E+05      & $10^{7}$ & 0.01 & 17.50 &6.5E+00\\

10 & 1 & 19.54 & 5.8E+04      & $10^{7}$ & 1 & 19.54 &7.0E+00\\

10 & 3 & 21.09 & 1.8E+02        & $10^{7}$ & 3 & 21.09 &7.0E+00\\

10 & 5 & 22.18 & 2.7E+01         & $10^{7}$ & 5 & 22.23 &6.4E+00\\

10 & 10 & 26.67 & 2.4E+00         & $10^{7}$ & 10 & 26.73 &1.9E+00\\

10 & 30 & 137.09 & 1.3E-01         & $10^{7}$ & 30 & 137.09 &1.3E-01\\

$10^{3}$ & 0.01 & 17.50 & 7.9E+03  & $10^{9}$ & 0.01 & 17.50 &3.5E-02\\

$10^{3}$ & 1 & 19.54 & 3.4E+04& $10^{9}$ & 1 & 19.54 &3.8E-02\\

$10^{3}$ & 3 & 21.09 & 1.8E+02  & $10^{9}$ & 3 & 21.09 &3.9E-02\\

$10^{3}$ & 5 & 22.18 & 2.7E+01   & $10^{9}$ & 5 & 22.28 &4.0E-02\\

$10^{3}$ & 10 & 26.67 & 2.4E+00   & $10^{9}$ & 10 & 27.10 &4.1E-02\\

$10^{3}$ & 30 & 137.09 & 1.3E-01   & $10^{9}$ & 30 & 143.88 &6.7E-02\\

$10^{5}$ & 0.01 & 17.50 & 3.2E+02 & $10^{11}$ & 0.01 & 17.50 &5.8E-05\\

$10^{5}$ & 1 & 19.54 & 5.4E+02  & $10^{11}$ & 1 & 19.54 &6.4E-05\\

$10^{5}$ & 3 & 21.09 & 1.6E+02  & $10^{11}$ & 3 & 21.09 &6.8E-05\\

$10^{5}$ & 5 & 22.23 & 2.7E+01   & $10^{11}$ & 5 & 22.28 &7.1E-05\\

$10^{5}$ & 10 & 26.67 & 2.4E+00   & $10^{11}$ & 10 & 27.10 &8.3E-05\\

$10^{5}$ & 30 & 137.09 & 1.3E-01   & $10^{11}$ & 30 & 151.36 &4.0E-04\\
\hline

\end{tabular}

\vspace*{5cm}
\end{table}
Table~\ref{tab1} shows the calculated $\beta^{+}$-decay rates on
$^{57}$Zn as a function of stellar temperature and density. In Table
1 the first column gives the stellar density in units of $gcm^{-3}$.
The second column gives the value of temperature in units of $10^{9}
K$. The third column gives the value of the calculated
$\beta^{+}$-decay rates in units of $sec^{-1}$ whereas the final
column shows the ratio of the calculated $\beta^{+}$-decay rates to
the electron capture rates for physical conditions given in first
two columns. One should note that even though terrestrially
$^{57}$Zn $\beta^{+}$-decays to $^{57}$Cu with a 100$\%$ ratio (no
electron capture), we do calculate a finite ratio of
$\beta^{+}$-decay to electron capture even at low temperature and
density (see Table~\ref{tab1}). This is because we calculate only
continuum electron capture and no bound states capture. It can be
seen from the table that there is no appreciable change in the
$\beta^{+}$-decay rates as the core stiffens from low density to
high density region. However the rates increase considerably with
increasing stellar temperature due to a considerable increase in the
available phase space.  At the later stages of the collapse,
$\beta^{+}$-decay becomes unimportant as an increased electron
chemical potential, which grows like $\rho ^{{1\mathord{\left/
{\vphantom {1 3}} \right. \kern-\nulldelimiterspace} 3} } $ during
in fall, drastically reduces the phase space. These results in
increased electron capture rates during the collapse phase. Electron
capture rates on $^{57}$Zn hence become important during the very
late phases of stellar evolution of massive stars (prior to
collapse) and at high stellar temperatures. At $\rho Y_{e}
[gcm^{-3}] = 10^{11}$ the calculated electron capture rates are
bigger than the competing $\beta^{+}$-decay rates by more than four
orders of magnitude.  For the relevant peak rp-process conditions
(T$_{9} [K] \sim 3$ and $\rho \sim 10^{7}$ gcm$^{-3}$), the pn-QRPA
calculated $\beta^{+}$-decay rates are around a factor seven bigger
than the corresponding electron capture rates. Therefore we present
our results for both $\beta^{+}$-decay rates of $^{57}$Zn and
electron capture rates on $^{57}$Zn and compare them with previous
calculations.

\begin{figure}[t]
\includegraphics[width=3.3in,height=4.3in]{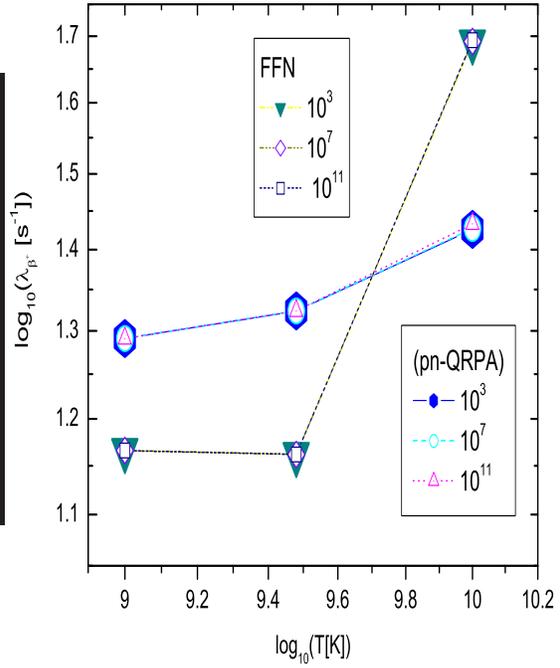} \caption {(
Color on line) Comparison of the pn-QRPA and FFN \cite{Fuller82}
$\beta^{+}$ decay rates of $^{57}$Zn nucleus as a function of
temperature for selected densities (in units of $g.cm^{-3}$ shown in
legends) in stellar matter.}\label{fig4}
\end{figure}
\begin{figure}[t]
\includegraphics[width=3.3in,height=4.3in]{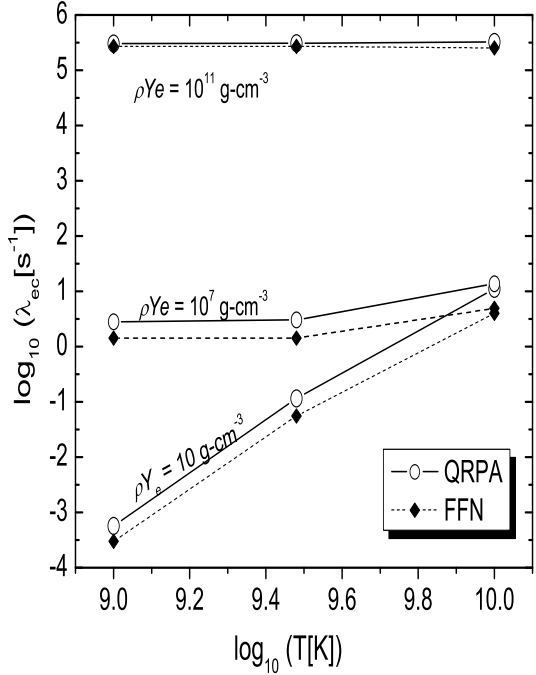} \caption
{Comparison of the pn-QRPA and FFN \cite{Fuller82} electron capture
rates on $^{57}$Zn nucleus as a function of temperature for selected
densities in stellar matter.}\label{fig5}
\end{figure}

To the best of author's knowledge, there are no shell model weak
rates available in the literature for $^{57}$Zn nucleus. Therefore,
pn-QRPA rates were compared with the pioneer work of Fuller et al.
\cite{Fuller82}. This comparison for the $\beta^{+}$ decay and
electron capture rates are shown in Fig.~\ref{fig4}, and
Fig.~\ref{fig5}, respectively. Fig.~\ref{fig4} shows that the
pn-QRPA $\beta^{+}$ decay rates are in good comparison with the FFN
rates. However at high stellar temperatures (T$_{9} [K] = 30$) the
FFN rates are almost twice the calculated rates. The authors in Ref.
\cite{Fuller82}  put no constraints on the threshold values of
parent excitation energies while in the present calculation such
constraints were incorporated through particle emission processes
\cite{Nabi04} which lead to the suppression of pn-QRPA $\beta^{+}$
decay rates at high stellar temperatures (when the probability of
occupation of high-lying parent excited states increases
substantially). Regarding the calculation of electron capture rates
(Fig.~\ref{fig5}) one notes that at high stellar densities (where
electron capture rates surpass the $\beta^{+}$ decay rates), $\rho
Y_{e} [gcm^{-3}] \sim 10^{11}$, the FFN rates are in good comparison
with the pn-QRPA electron capture rates. At lower densities the
pn-QRPA electron capture rates are bigger by a factor 2 (T$_{9} [K]
\sim 1$) to a factor of 3 when T$_{9} [K] = 30$. This enhancement is
attributed to the low placement of the GT centroid in the daughter
nucleus, $^{57}$Cu, by the pn-QRPA model.

The $\beta^{+}$-decay and electron capture rates on $^{57}$Zn were
calculated on a fine grid temperature-density scale suitable for
simulation codes and for necessary interpolation purposes. The ASCII
files of these rates may be requested from the corresponding author.

\section{Summary}
The GT strength function is an important ingredient in the complex
dynamics of presupernova and supernova explosion since these GT
transitions partly determine the $\beta^{+}$-decay and electron
capture rates in the stellar core. The GT strength distributions in
$^{57}$Zn were calculated within the domain of the pn-QRPA theory.
The pn-QRPA calculated GT strength showed differences with the
earlier reported shell model calculation. The pn-QRPA results for
$^{57}$Zn were in good agreement with the experimental results
\cite{Jokinen02} and reproduced well the measured GT strength
between 2-3 MeV. The stellar weak decay rates are sensitive to the
location and structure of these low-lying states in daughter
$^{57}$Cu. Small changes in the binding and excitation energies can
lead to significant modifications of the predictions for the
synthesis of proton-rich isotopes. The primary mechanism for the
production of the proton-rich nuclei is the rp-process and is
believed to be important in the dynamics of the collapsing
supermassive stars. The good agreement of the reported low-lying GT
strength with the experimental results validates the choice of the
pn-QRPA theory as a preferred model for the calculation of weak
rates for proton-rich nuclei. The choice of nuclear model can affect
the prediction and synthesis of proton-rich isotopes as well as the
time evolution of the cosmic x-ray burst and dynamics of the
collapsing supermassive stars. For comparison, in the present work
we obtained a summed B(GT) of 1.86 as compared to 1.25 in Ref.
\cite{Jokinen02} within the excitation energy interval between 0 and
7 MeV in daughter $^{57}$Cu. The shell model predicted strength for
the same excitation energies amounts to 1.68.

The calculated GT strength functions were further used to calculate
electron capture and $\beta^{+}$-decay rates of $^{57}$Zn in stellar
matter, particularly for rp-process conditions. At high stellar
temperatures the calculated $\beta^{+}$-decay rates on $^{57}$Zn are
half of the corresponding FFN calculated rates. For typical peak
rp-process conditions, T$_{9} [K] \sim 3$ and $\rho \sim 10^{7}$
gcm$^{-3}$, the pn-QRPA calculated $\beta^{+}$-decay (electron
capture) rates on $^{57}$Zn are bigger by a factor of 1.5 (2). This
may have interesting astrophysical consequences for collapse
simulators.

Realistically speaking weak interaction mediated rates of hundreds
of nuclei are involved in the complex dynamics of supernova
explosion . Incidently, the most abundant nuclei tend to have small
weak rates as they are more stable and the most reactive nuclei tend
to be present in minor quantities. Thus, the most important in the
stellar core is the rate times abundance of a particular specie. We
are in the process to calculate microscopically the weak decay rates
for nuclei which are considered to be important in astrophysical
environment as part of this on-going project. Few of such important
weak rates were recently presented (e.g.
\cite{Nabi08a,Nabi08b,Nabi09,Nabi09a,Nabi10,Nabi10a}). Work is still
in progress for the microscopic calculation of weak rates of
remaining key iron-regime nuclei. Core-collapse simulators are urged
to test run the pn-QRPA reported weak rates in typical stellar
conditions to check for probable interesting outcomes.

\end{document}